\documentclass[fleqn,usenatbib]{mnras}

\usepackage{newtxtext,newtxmath}

\usepackage[T1]{fontenc}

\DeclareRobustCommand{\VAN}[3]{#2}
\let\VANthebibliography\thebibliography
\def\thebibliography{\DeclareRobustCommand{\VAN}[3]{##3}\VANthebibliography}

\usepackage{graphicx}	
\usepackage{amsmath}	
\usepackage{CJK}

\title[CDGs in SDSS]{Discovery of compact disc Galaxies with High Surface Brightness in the Sloan Digital Sky Survey}

\author[Chen \& Hwang]{
Cheng-Yu Chen (陳振予)$^{1}$\thanks{E-mail: m969004@gm.astro.ncu.edu.tw}
and Chorng-Yuan Hwang (黃崇源)$^{1}$
\\
$^{1}$Graduate Institute of Astronomy, National Central University, Taoyuan City 320317, Taiwan (R.O.C.)\\
}

\date{Accepted 2023 August 28. Received 2023 July 28; in original form 2022 December 21}
\pubyear{2022}

\begin{document}
\begin{CJK*}{UTF8}{bkai}
\label{firstpage}
\pagerange{\pageref{firstpage}--\pageref{lastpage}}
\maketitle
\end{CJK*}

\begin{abstract}
Compact disc galaxies (CDGs) with high surface brightness were identified in the Sloan Digital Sky Survey (SDSS) data.
We determined the surface profiles of the CDGs and compared them to those of normal-sized disk galaxies (NDGs).
The CDGs have higher central brightness and older stellar age than the NDGs.
Furthermore, the brightness profiles of the CDGs fit a S{\'e}rsic model with $n \approx 2.11$ and have a zero $g^{\prime}-r^{\prime}$ color gradient on average.
By contrast, the NDGs fit an exponential profile and have a negative color gradient on average.
These results indicate that the structure and stellar population of the CDGs and NDGs differ.
We suggest that the CDGs are ancient galaxies in the quenching phase following the initial central starburst.
\end{abstract}




\section{Introduction}

One key feature of a galaxy is its brightness.
\citet{1970ApJ...160..811F} reported that disk galaxies have similar central surface brightness ($21.65\pm0.3$ B-mag per squared arcsec).
\citet{2010ApJ...722L.120F} derived the central surface brightness of galaxies in Sloan Digital Sky Survey (SDSS) images and obtained an average value of $\langle \mu_0 \rangle =20.2\pm0.7~\mathrm{mag}~\mathrm{arcsec^{-2}}$ in the $r-$band.
However, several studies have reported a population with surface brightness significantly lower than this average value.
These objects are known as low-surface-brightness galaxies \citep[LSBGs, e.g.][]{2003A&A...405...99M,2011ApJ...728...74G,2019MNRAS.483.1754D}.
However, galaxies that are brighter than this average have rarely been studied.
Such galaxies would have a central brightness noticeably greater than those of typical galaxies.

The brightness of a galaxy is related to its mass and stellar population; several studies have reported a relationship between galaxy mass and size.
On a logarithmic scale, the stellar mass of a spiral galaxy and its effective radius have either a linear \citep[e.g.][]{2010MNRAS.406.1595F,2018MNRAS.473.5468W} or quadratic \citep{2013MNRAS.434..325F} correlation.
However, early-type galaxies have a different relation between stellar mass and effective radius than late-type galaxies \citep{2013MNRAS.434..325F,2014ApJ...788...28V}.
Most disc galaxies have a mass--size relation clustered around $R \sim M^\alpha$, $\alpha = 0.26-0.38$ \citep[e.g.][]{2010ApJ...722L.120F,2018MNRAS.473.5468W}.
However, this relationship does not hold for some galaxies.
These galaxies have large stellar masses but significantly smaller effective radii than those of typical galaxies with similar masses.
Galaxies in this population are known as massive compact galaxies (MCGs).

MCGs have been investigated by several studies.
\citet{2012MNRAS.423..632F} studied MCGs from the NYU value-added Galaxy Catalogue \citep{2005AJ....129.2562B}.
They observed that local MCGs are fast rotators with elongated morphologies and have young luminosity-weighted ages with high metallicities.
They suggested that local MCGs might be formed from starbursts triggered by gas-rich mergers.
\citet{2015ApJ...813...23V} studied the mechanism underlying the formation of MCGs from the 3D-HST project \citep{2011ApJ...743L..15V,2012ApJS..200...13B}.
They suggested that in MCG evolvution, the mass of the galaxy increases while its radius remains fixed.
\citet{2021MNRAS.507..300S} studied MCGs in the Mapping Nearby Galaxies project at Apache Point Observatory \citep[MaNGA,][]{2015ApJ...798....7B}.
Their results demonstrated that the MCGs are more metal-rich and $\alpha$-enhanced than typical galaxies, and they suggested that MCGs are the descendants of compact post starburst galaxies.
Most MCGs described in previous studies have a stellar mass greater than $10^{11}~\mathrm{M}_{\sun}$ \citep[e.g.][]{2012ApJS..200...13B,2015ApJ...813...23V}.
Only a few researchers have studied compact galaxies with a stellar mass between $10^{10}$ and $10^{11}~\mathrm{M_{\sun}}$ \citep[][]{2021MNRAS.507..300S}, and the properties of compact galaxies with a stellar mass less than $10^{10}~\mathrm{M_{\sun}}$ are unknown.

In this study, we selected compact disk galaxies from SDSS that deviate from the typical mass--size distribution.
We compared the brightness, stellar age, and star formation history of these galaxies with those of the normal size galaxies of similar stellar masses.
Our data selection method is described in Section 2.
In Section 3, we derive the surface brightness profiles of our sample.
In Section 4, we discuss the photometric and stellar properties of our sample; Section 5 summarizes and concludes our work.

\section{Data Selection}

We selected galaxies from the 16th data release (DR16) of SDSS \citep[][]{2020ApJS..249....3A}.
We first selected galaxies with a spectroscopic redshift between 0.01 and 0.08.
Second, we classified the morphology of the galaxies on the basis of data from the Galaxy Zoo project \citep{2008MNRAS.389.1179L,2011MNRAS.410..166L}.
Galaxies with a fraction of votes for spiral greater than 0.7 were considered spiral galaxies.
In the SDSS database, galaxies were fitted to a linear combination of an exponential profile and a de Vaucouleurs profile as follows:
\begin{equation}
F_{composite} = fracDeV F_{deV} + (1-fracDeV)F_{exp}.
\end{equation}
We then selected galaxies with a $fracDeV$ of zero; the behavior of these galaxies fit well to a pure exponential disk.
A total of 24,097 such spiral galaxies with a $fracDeV$ of zero were identified.
We expected that these galaxies would have a simple mass--size relationship. 
The stellar mass of each galaxy was derived by applying the 2016 updated version of the \citet{2003MNRAS.344.1000B} stellar population synthesis (SPS) model, the \citet{2003PASP..115..763C} initial mass function, and the Padova 1994 evolution track (see Section 2.1 of \citet{2003MNRAS.344.1000B} in details).
This evolution track consists of seven metallicities with $Z=0.0001,0.0004,0.004,0.008,0.02,0.05,0.1$.
The \citet{2003MNRAS.344.1000B} models consist of 221 stellar age bins spanning from 10~Myrs to 20~Gyrs.
We truncated the model age at 13.8~Gyrs for our analysis.
We adopted the spectral redshifts and the foreground extinctions from SDSS to convert the model spectra into the apparent magnitude in the five SDSS filters.
For each galaxy, we compared the converting magnitudes to the observed magnitudes by applying the least squares method to obtain the optimal model spectrum.
The mass-to-luminosity ratio in the $r^{\prime}$-band obtained from the optimal model was then adopted.
The cosmological parameters adopted for our analysis are $H_0 = 70 ~\mathrm{km~s^{-1}~Mpc^{-1}}$, $\Omega_M = 0.3$, and $\Omega_{\Lambda} = 0.7$.

\begin{figure}
        \includegraphics[width=\columnwidth]{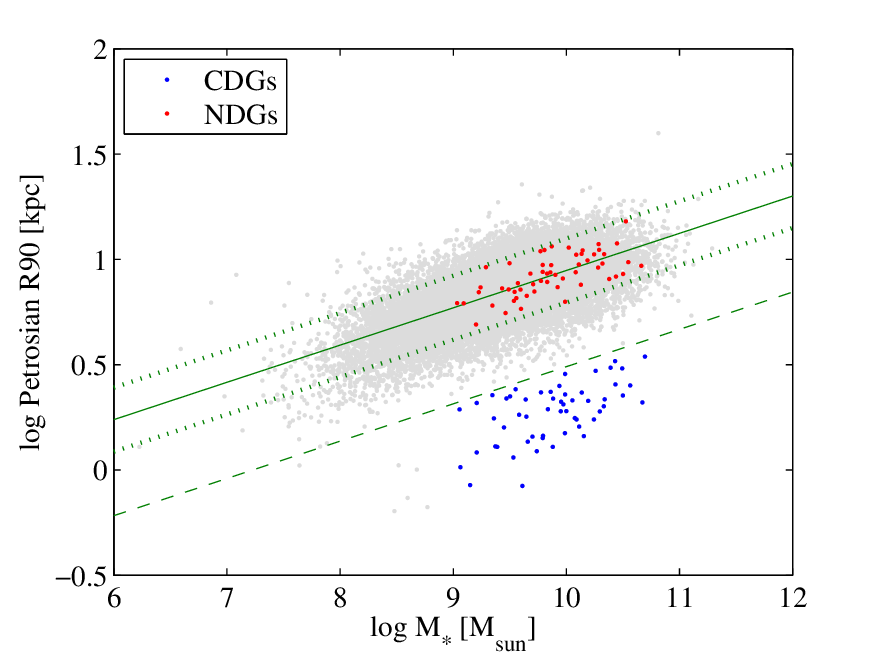}
    \caption{SDSS $R_{90}$ as a function of stellar mass. Red dots denote the NDGs. The solid and dashed lines represent the fitted size-mass relation and the fitted line minus $3\sigma$, respectively. The dotted lines represent one sigma around the fitted line. The CDGs were selected based on the galaxy size below the dashed line and the stellar mass greater than $10^{9}~\mathrm{M_{\sun}}$.}
    \label{fig1}
\end{figure}
Fig. \ref{fig1} displays the relationship between stellar mass $m_*$ and the radius containing $90\%$ of the Petrosian flux ($R_{90}$).
Most of the galaxies fall along a linear distribution.
Hence, we fitted this relationship as follows.
\begin{equation}
\log R_{90} = a\log m_* + b,
\end{equation}
where $R_{90}$ is in kiloparsecs, $m_*$ is in solar masses, $a=0.177$, and $b=0.823$.
The solid line in Fig. \ref{fig1} represents the fitted trendline.
We then fit the distribution of $\Delta y = \log R_{90} - a\log m_* - b$ to a Gaussian distribution.
The variance $\sigma_{\Delta y}$ of the fitted Gaussian distribution is 0.152. 
The dashed line in Fig. \ref{fig1} is the solid line minus $3\sigma_{\Delta y}$, and the dotted lines represent one sigma around the fitted line.
Galaxies that fall within the dotted lines were named normal-sized disk galaxies (NDGs); those below the dashed line are significantly smaller than the NDGs but have similar stellar masses.
These smaller galaxies below the dashed line were named compact disk galaxies (CDGs).
For our CDG sample, we selected galaxies with a mass greater than $10^{9}~\mathrm{M_{\sun}}$ below the dashed line, resulting in a total of 56 galaxies. 
Their data, along with other results, are summarized in Table \ref{tab:table_data}.
For comparison, we also selected 56 NDGs within the mass range of $\log M / M_{\odot} = 9$ to $\log M / M_{\odot}=11$, divided into eight bins with a bin size of 0.25.
In each mass bin, we randomly selected the same number of galaxies as the SDGs within 1-$\sigma_{\Delta y}$ of the main mass-size relation to form our NDG sample.
The NDGs are denoted as red points in Figure \ref{fig1}.
The selected NDGs have stellar masses similar to those of the CDGs.
We then compare the surface brightness and stellar populations of these NDG and CDG samples.

\section{Results}
We measured the surface brightness profiles of our samples by analyzing the SDSS images in all filters.
We wrote Matlab programs to measure the profiles. 
For each galaxy, we first fitted a Gaussian to the distribution of all pixel values within four times the Petrosian radius by applying the fitting function of Matlab.
If the flux of a pixel was from a source, such as a star or a galaxy, the flux should be much greater than the sky level.
Therefore when we fit the pixel values to a Gaussian distribution, we will get a Gaussian with a long tail.
The fitted mean and variance were applied to subtract the sky background.
Next, we identified all neighbors within four times the Petrosian radius of the source.
For each neighbor, we removed its flux by excluding the pixel values within an ellipse with a major axis of two times the Petrosian radius for a neighbor galaxy and within a circle with a radius of two times the Petrosian radius for a neighbor star.
We then divided each galaxy into many concentric elliptical rings with two-pixel wide using the central points, position angles, and axis ratios obtained from SDSS.
We used the average flux within various elliptical rings for each galaxy to represent the surface brightness profile.
The intensity profile of each galaxy was truncated at $3\sigma$ of the sky level.
The sum of the flux within the truncated radius was considered the total flux $F_{tot}$.
We then derived the effective radius $R_\mathrm{eff}$ and 90\%-flux radius $R_{90}$ based on the $F_{tot}$ values.

\begin{figure}
        \includegraphics[width=\columnwidth]{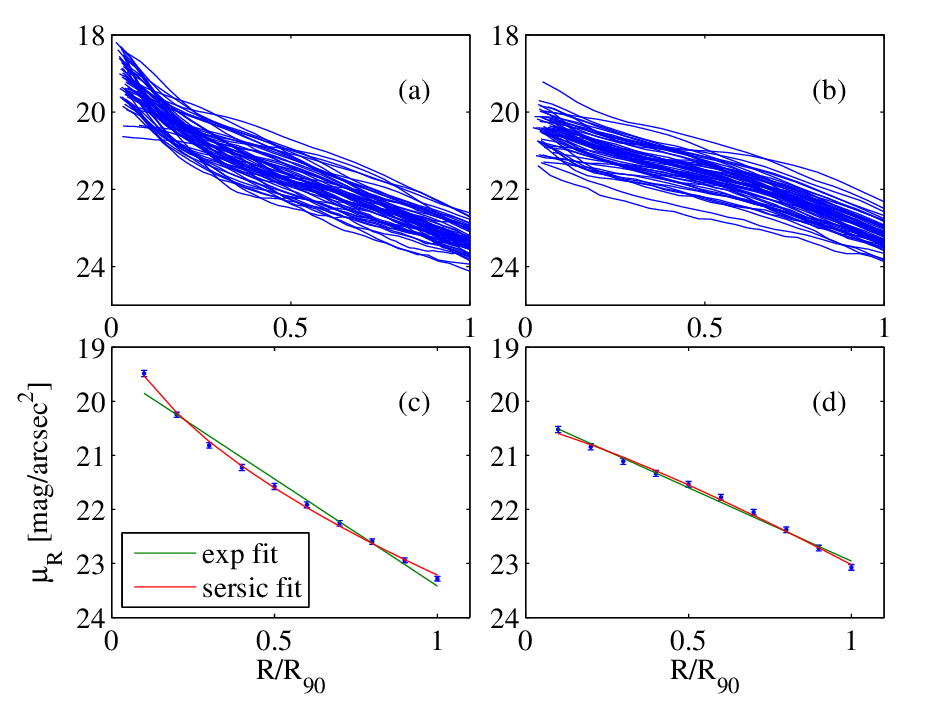}
    \caption{Surface brightness profiles in the $r^{\prime}-$band of (a) CDGs and (b) NDGs within $R_{90}$. The dots in (c) and (d) indicate the average profiles of the galaxies in (a) and (b), respectively. The green and red lines in (c) and (d) are the fitted exponential and S{\'e}rsic models, respectively, for the average profiles.}
    \label{fig2}
\end{figure}

Figure \ref{fig2} presents the $r^{\prime}-$band surface brightness profiles of the samples.
For each galaxy, we fitted the intensity profile to a S{\'e}rsic model.
In a S{\'e}rsic model, the surface brightness can be expressed as a function of radius $r$ \citep[][]{2005PASA...22..118G} 
\begin{equation}
\label{eq3}
    \mu(r) = \mu_0 + \frac{2.5}{\ln{10}} (\frac{r}{r_d})^{(1/n)},
\end{equation}
where $\mu_0$ is the central surface brightness, $r_d$ is the disk scale length, and $n$ is the S{\'e}rsic index.
An exponential profile is a S{\'e}rsic model with $n=1$.
We fitted the intensity profile of each galaxy to equation~(\ref{eq3}) both using $n=1$ and with $n$ as a free parameter.
The selected galaxies have radii greater than 5.5 arcsecs, which are much greater than the resolution of SDSS.
Therefore, we did not consider the effects of the point spread function.
Because we selected galaxies with a $fracDeV$ of 0, the galaxies were expected to have an exponential profile.
Figure \ref{fig2}(c) and (d) present the fitted exponential and S{\'e}rsic models for the average profiles of the CDGs and the NDGs.
The results indicate that the NDGs have nearly perfect exponential profiles and also fit exponential profiles with S{\'e}rsic indices $\approx$1.
By contrast, the CDG profiles differ greatly from an exponential profile; the fitted S{\'e}rsic indices are 2.11 on average.
Typically, the CDG profile decreases rapidly in the central region and stabilizes in the outer region.
These results indicate that stars in the CDGs are more centrally concentrated than in the NDGs.

We derived three brightness values from the exponential fit of the surface brightness profile, the average flux within the effective radius $R_e$, and the average flux of the fiber magnitude from SDSS.
The three brightness values are denoted as $\mu_0$, $\mu_e$, and $\mu_\mathrm{fib}$, respectively.
The central brightness values of the exponential fit were obtained using equation~(\ref{eq3}) with $n=1$.
The average flux within the effective radius $R_\mathrm{eff}$ was obtained from the SDSS images, and the average flux within the SDSS 3-arcsec-diameter fibers was directly obtained from the SDSS fiber flux.

\begin{table}
        \centering
        \caption{Mean and variance of the central surface brightness in the $r^{\prime}$-band.}
        \label{tab:table1}
        \begin{tabular}{lcccccc} 
                \hline
                & \multicolumn{2}{c}{Exponential fit} & \multicolumn{2}{c}{Average in $R_\mathrm{eff}$} & \multicolumn{2}{c}{Average in fiber }\\
                & $\langle \mu_0 \rangle$ & $\sigma_{\mu_0}$ & $\langle \mu_e \rangle$ & $\sigma_{\mu_e}$ & $\langle \mu_\mathrm{fib} \rangle$ & $\sigma_{\mu_\mathrm{fib}}$ \\
                \hline
                CDGs & 19.59 & 0.39 & 20.85 & 0.41 & 19.78 & 0.50\\
                NDGs & 20.31 & 0.47 & 21.33 & 0.43 & 20.85 & 0.40\\
                \hline
        \end{tabular}
\end{table}

\begin{table}
        \centering
        \caption{K--S test results for surface brightness.}
        \label{tab:table2}
        \begin{tabular}{lcccc} 
                \hline
                & Exponential fit & Average in $R_\mathrm{eff}$ & Average in fiber\\
                \hline
                $p$ & 7.12 $\times 10^{-10}$ & 5.97 $\times 10^{-5}$ & $<\times 10^{-10}$ \\
                \hline
        \end{tabular}
\end{table}

\begin{figure}
        \includegraphics[width=\columnwidth]{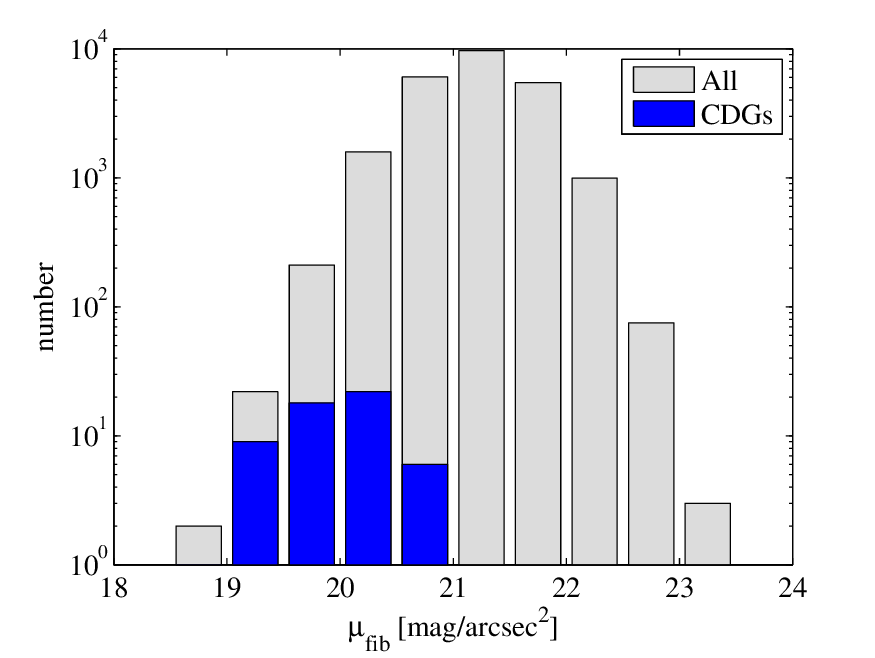}
    \caption{Surface brightness distributions of our samples.
    Blue and grey bars represent the CDGs and all spirals with $fracDeV=0$, respectively.
    CDGs dominate the results on the bright side.}
    \label{fig3}
\end{figure}

Figure \ref{fig3} presents the average brightness distributions of the CDGs and NDGs derived from the SDSS data on the basis of the average flux within the fiber radius.
The CDGs had higher central brightness values than typical spirals do.
Table \ref{tab:table1} presents the mean and variance values of the surface brightness of our samples.
We compared the distribution of the central brightness between the CDGs and the NDGs by using the Kolmogorov-Smirnov (K--S) test statistics.
The $p$ values of the K--S test are listed in Table \ref{tab:table2}.
We observed that the central regions of the CDGs were significantly brighter than those of the NDGs.

\begin{figure}
        \includegraphics[width=\columnwidth]{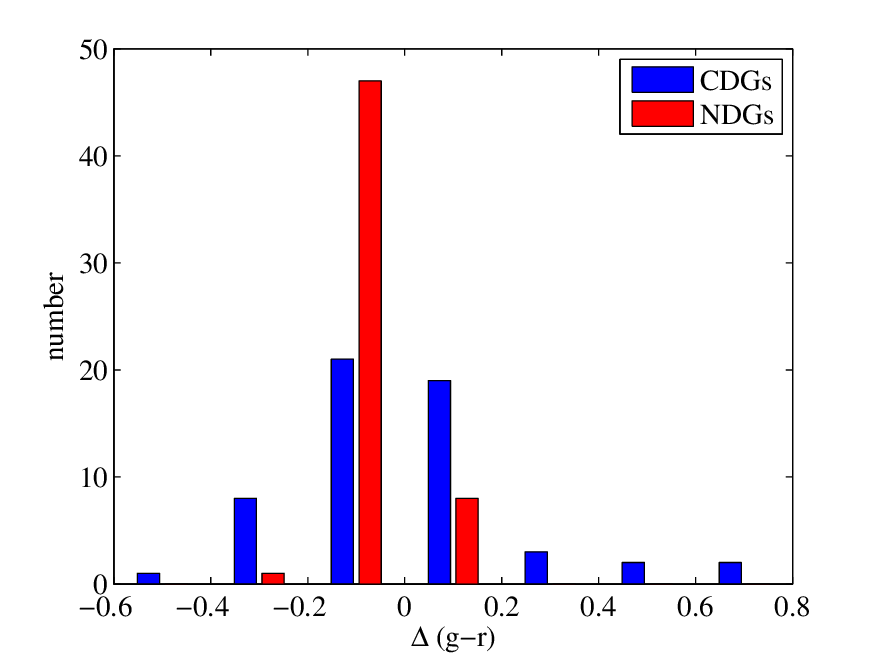}
    \caption{CDG and NDG color gradient distributions.
    Blue and Red bars represent the CDGs and NDGs, respectively.
    CDGs have a greater average color gradient value than NDGs.}
    \label{fig4}
\end{figure}

We investigated the color gradients of our selected sample.
For each galaxy, we applied the $R_\mathrm{eff}$ and the $3\sigma$ edge of the $r^{\prime}-$band image to the brightness profiles of all filters to measure the magnitudes inside and outside the $r^{\prime}-$band $R_\mathrm{eff}$.
The $g^{\prime}-r^{\prime}$ color gradient distributions for our samples are displayed in Figure \ref{fig4}.
Table \ref{tab:table3} lists the mean and variance values of the color gradients.
We compared the K--S test statistics for the CDG and NDG color gradient distributions.
The $p$ values of the K--S test is $4.97 \times 10^{-4}$.
The color gradient distribution of the CDGs shows a significant difference from the color gradient distribution of the NDGs.
The CDGs typically have zero color gradients, whereas the NDGs have negative color gradients.

\begin{table}
        \centering
        \caption{Mean and variance values of the $g-r$ color gradients.}
        \label{tab:table3}
        \begin{tabular}{cccc} 
                \hline
                & $\langle \Delta_{g-r} \rangle$ & $\sigma_{\Delta_{g-r}}$ \\
                \hline
                CDGs & 0.0087 & 0.24 \\
                NDGs & -0.082 & 0.080 \\
                \hline
        \end{tabular}
\end{table}

\begin{figure}
        \includegraphics[width=\columnwidth]{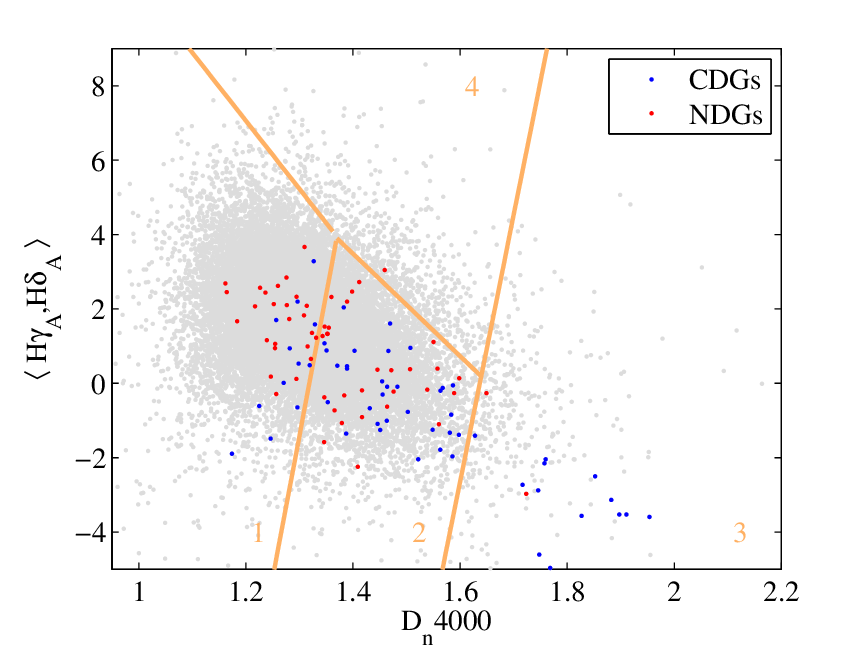}
    \caption{Balmer absorption indices and $4000~\text{\AA}$ breaks of our samples.
    Blue and Red dots represent the CDG and NDG samples, respectively, and grey dots represent the total sample of spiral galaxies with $fracDeV=0$. Orange numbers denote the four regions of the model of \citet{2015MNRAS.451..433H}.}
    \label{fig5}
\end{figure}

\begin{figure}
        \includegraphics[width=\columnwidth]{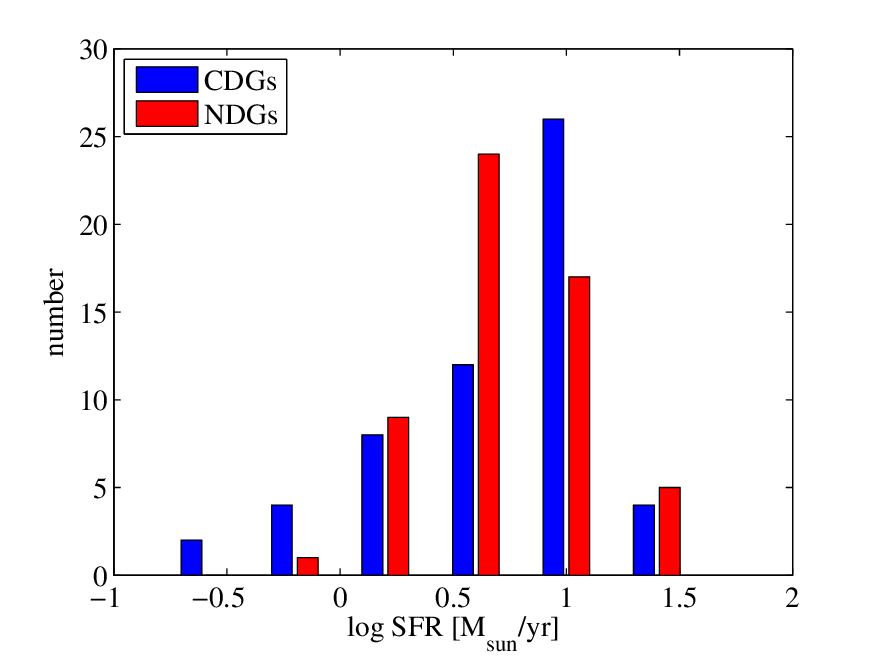}
    \caption{Star formation distributions of our samples.
    Blue and Red bars represent the CDGs and NDGs, respectively.
    The CDG SFR distribution has a tail on the low-SFR side.}
    \label{fig6}
\end{figure}

We investigated the star formation history of our sample by applying $\mathrm{H\delta}$ and $\mathrm{H\gamma}$ absorption indices and $\mathrm{D_n 4000}$ from SDSS \citep[][]{2004MNRAS.351.1151B} and comparing the results with the star formation model of \citet{2015MNRAS.451..433H}.
The model of \citet{2015MNRAS.451..433H} simulates the evolution paths of different star formation rates.
Galaxies with different star formation history would follow different evolution tracks in the $\langle \mathrm{H\gamma_A,H\delta_A}\rangle$ -- $D_n 4000$ diagram.
Figure \ref{fig5} presents the average Balmer absorption index $\langle \mathrm{H\gamma_A,H\delta_A}\rangle$ as a function of $D_n 4000$.
Regions 1--4 are the same as those in the model of \citet{2015MNRAS.451..433H}.
If a galaxy underwent a starburst within the last 2~Gyr, the evolution path begins from region 1 and goes to region 3 through region 4.
Otherwise, the evolution path begins from region 1 and goes to region 3 through region 2.
The CDGs are all in regions 1, 2, and 3, indicating that starbursts did not occur in these galaxies within the last 2~Gyr.
We further investigated the star formation rates (SFRs) of our samples from the SDSS data \citep{2003MNRAS.341...33K}.
Figure \ref{fig6} presents the SFR distributions.
We compared the CDG and NDG SFR distributions by using the K--S test statistics.
The $p$ values of the K--S test are 0.30, which suggests that there are no significant differences in the SFRs between the CDGs and NDGs.

\begin{figure}
        \includegraphics[width=\columnwidth]{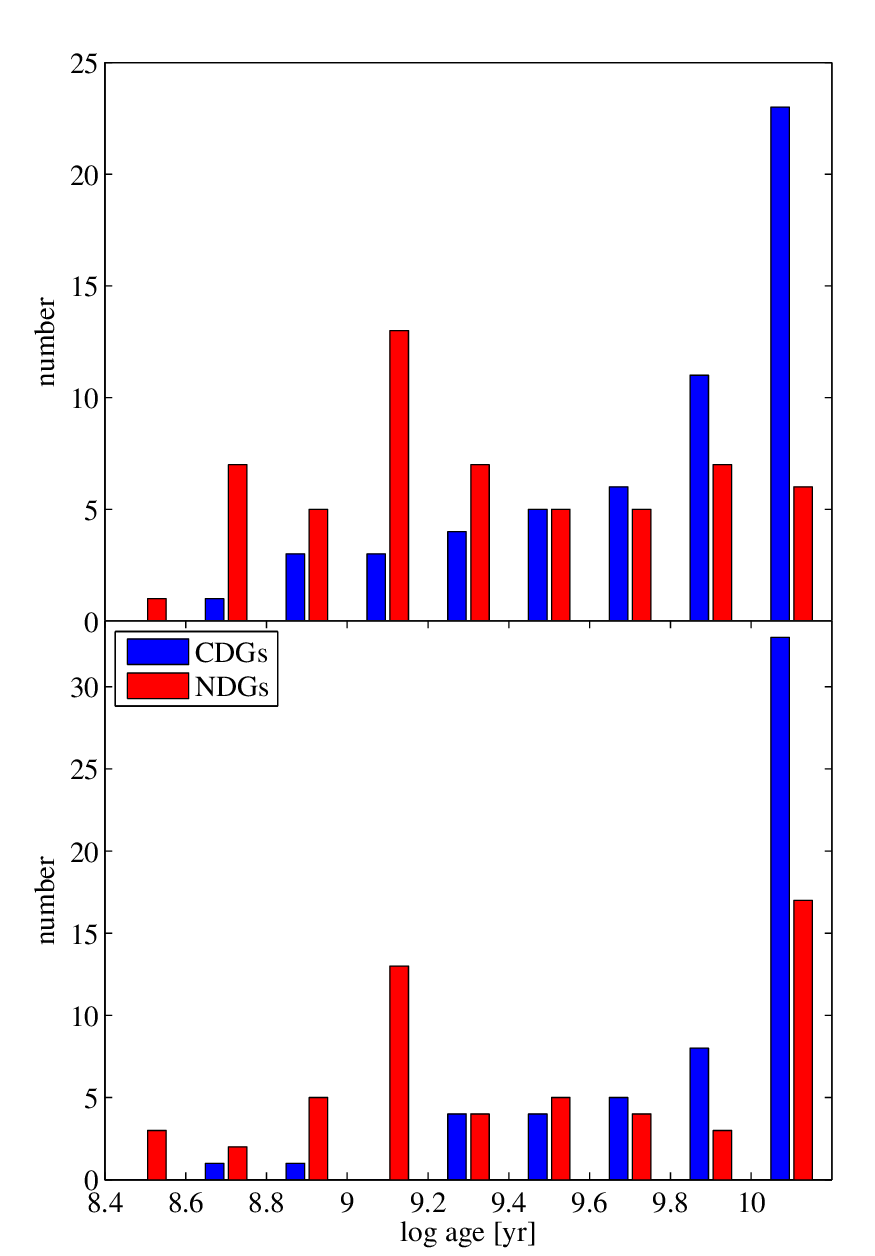}
    \caption{Fitted stellar age distributions of our galaxy samples.
    Top: Inside $R_\mathrm{eff}$. Bottom: Outside $R_\mathrm{eff}$.
    Blue and Red bars represent the CDGs and NDGs, respectively.}
    \label{fig7}
\end{figure}

We investigated the stellar ages of the regions inside and outside the $R_\mathrm{eff}$ of our samples.
We applied the same methods as we estimated the stellar mass to derive the optimal spectra in the \citet{2003MNRAS.344.1000B} models with the same metalliticities and adopted the stellar ages of these spectra.
Figure \ref{fig7} displays the stellar age distributions of our samples.
The mean and variance values of the ages are listed in Table \ref{tab:table4}.
We compared the CDG and NDG stellar age distribution by using the K--S test statistics.
The $p$ values of the K--S are listed in Table \ref{tab:table5}.
The results indicate that the CDGs have significantly older stellar populations than the NDGs both inside and outside the effective radius.

\begin{table}
        \centering
        \caption{Mean and variance of the fitted stellar ages.}
        \label{tab:table4}
        \begin{tabular}{ccccc} 
                \hline
                & \multicolumn{2}{c}{$R<R_\mathrm{eff}$} & \multicolumn{2}{c}{$R>R_\mathrm{eff}$}\\
                & $\langle \log age \rangle$ & $\sigma_{\log age}$ & $\langle \log age \rangle$ & $\sigma_{\log age}$ \\
                & [yr] & [yr] & [yr] & [yr] \\
                \hline
                CDGs & 9.77 & 0.42 & 9.90 & 0.35 \\
                NDGs & 9.34 & 0.47 & 9.48 & 0.54 \\
                \hline
        \end{tabular}
\end{table}

\begin{table}
        \centering
        \caption{K--S results for the fitted stellar ages.}
        \label{tab:table5}
        \begin{tabular}{lcc} 
                \hline
                & $R<R_\mathrm{eff}$ & $R>R_\mathrm{eff}$\\
                \hline
                $p$ & 3.92 $\times 10^{-5}$ & 2.21 $\times 10^{-4}$ \\
                \hline
        \end{tabular}
\end{table}

\section{Discussion}

We selected galaxies of various sizes but in the same stellar mass range and determined their surface brightness.
The selected galaxies all had $fracDeV=0$ in SDSS.
The surface brightness profiles were expected to approximate the exponential profile or a S{\'e}rsic profile with an index of approximately one, and this was the case for the NDGs.
By contrast, the CDGs did not fit the exponential profile well, and their fitted S{\'e}rsic indices were $2.11\pm0.18$ on average.
We also derived the $R_\mathrm{eff}/R_\mathrm{90}$ ratios of the galaxies; these were $0.47\pm0.05$ and $0.54\pm0.03$ for the CDGs and NDGs.
The K--S test result revealed a very small $p$-value of less than $10^{-10}$, indicating that the CDGs are significantly more concentrated than the NDGs.
Table \ref{tab:table1} shows that the average $\mu_0$ value of the NDGs obtained from the exponential fit is $20.31\pm0.47~\mathrm{mag}~\mathrm{arcsec^{-2}}$.
\citet{2010ApJ...722L.120F} fitted spiral galaxies to an exponential disk and obtained $\langle \mu_0 \rangle =20.2\pm0.7~\mathrm{mag}~\mathrm{arcsec^{-2}}$.
We compared the central surface brightness values of our NDGs with those of the \citet{2010ApJ...722L.120F} galaxies using a Student's t-test, which yielded a $p$-value of 0.11.
In other words, the exponential fit of the NDGs resulted in an average $\mu_0$ value that is consistent with the value reported by \citet{2010ApJ...722L.120F}.
Therefore, the CDGs have higher central surface brightness values than typical spiral galaxies.

We fit the entire profiles to an exponential model instead of extrapolating disk fits into the center as in previous studies \citep{1970ApJ...160..811F,2010ApJ...722L.120F}.
It is obvious from Figure \ref{fig2} that if we only fit the disk and extrapolate to the center to derive the central brightness, we will underestimate the true central brightness for the CDGs.
For the NDGs, the method used by \citet{1970ApJ...160..811F} and \citet{2010ApJ...722L.120F} seems reasonable because the NDGs are consistent with disk profiles as shown in Figure \ref{fig2}(b).
This suggests that there is an additional component for the CDGs.

The CDGs are relatively small for galaxies of their mass.
\citet{2021MNRAS.507..300S} selected the $10\%$ of galaxies with the smallest $R_\mathrm{eff}$in each mass bin as MCGs.
Although we attempted to select small galaxies in these mass ranges, our CDGs are not as small as the MCGs of \citet{2021MNRAS.507..300S}.
The $R_\mathrm{eff}$ of our CDGs are $5.37~\mathrm{kpc}$ on average, whereas the $R_\mathrm{eff}$ of the MCGs are <$1.5~\mathrm{kpc}$ \citep[e.g.,][]{2012MNRAS.423..632F,2015ApJ...813...23V,2021MNRAS.507..300S}.
Hence, the CDGs are less compact than the MCGs.
We suggest that the CDGs belong to a different population than the MCGs.
One possible formation mechanism of the MCGs is a central gas-rich starburst \citep[][]{2012MNRAS.423..632F,2021MNRAS.507..300S}.
\citet{2018ApJ...868...37W} suggested an evolution path in which galactic size $R_\mathrm{eff}$ decreases after a central starburst.
However, Figure \ref{fig5} indicates that our CDGs did not experience a starburst within 2~Gyrs.
Furthermore, Figure \ref{fig6} reveals that the average SFRs of the CDGs and of the NDGs are indistinguishable and that the SFR distribution of the CDGs has a low-SFR tail.
We suggest that the CDGs are galaxies in the quenching phase of star formation.

The CDGs have older stellar populations than the NDGs, indicating that their stellar populations formed earlier than those of the NDGs.
Because the star formation histories of these galaxies with different mass--size relationships are similar, we contend that the initial starbursts of the CDGs were more powerful than those of the NDGs and that the CDGs are at a later evolutionary stage. 
We suggest that the progenitors of the CDGs evolved through the fast process reported by \citet{2018ApJ...868...37W}; they initially experienced a central starburst at the beginning and became highly compact objects.
After this starburst, they evolved through the slow process of \citet{2018ApJ...868...37W}; they slowly increased in mass and size and ultimately became the small galaxies identified in this study.
Therefore, the CDGs are most likely relics of ancient compact objects that did not undergo a recent central starburst.

\begin{figure}
       \includegraphics[width=\columnwidth]{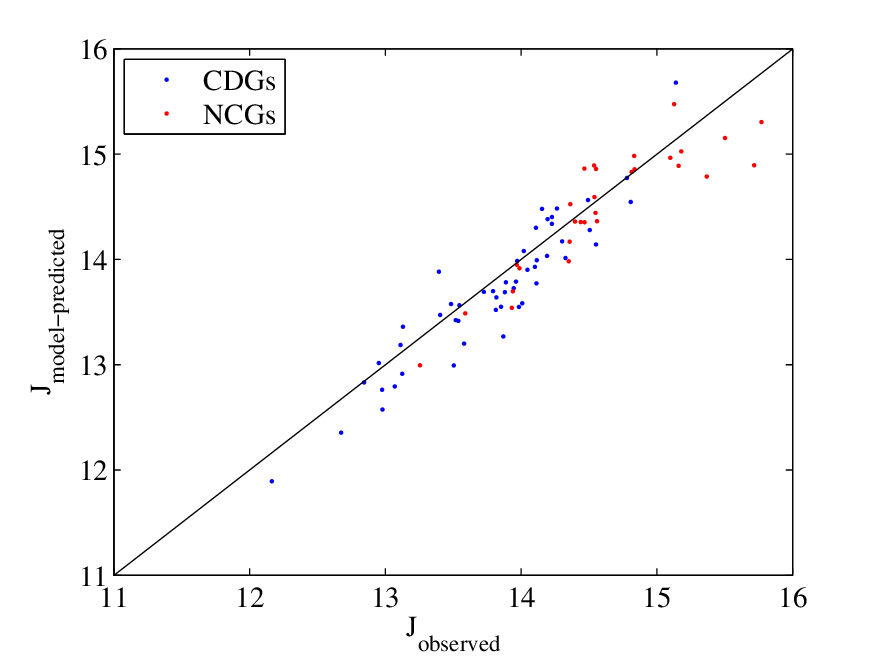}
    \caption{
    Observed and model-predicted 2MASS J-band magnitude.
    Blue and Red points represent the CDGs and NDGs, respectively.
    }
    \label{fig8}
\end{figure}

We did not take into account the effect of dust reddening in our estimation of stellar ages, which could potentially affect the accuracy of our results.
In order to evaluate the potential impact of dust, we compared the model-predicted values of \citet{2003MNRAS.344.1000B} to the observed values of the Two Micron All-Sky Survey (2MASS) J-band magnitudes for our selected galaxies.
Figure \ref{fig8} displays the observed and model-predicted 2MASS J-band magnitudes for the 51 CDGs and 29 NDGs that have available J-band magnitudes from the SDSS TwoMassXSC table.
The results indicate that the model-predicted values are consistent with the observed values, suggesting that dust reddening did not have a significant impact on our findings.

\begin{figure}
       \includegraphics[width=\columnwidth]{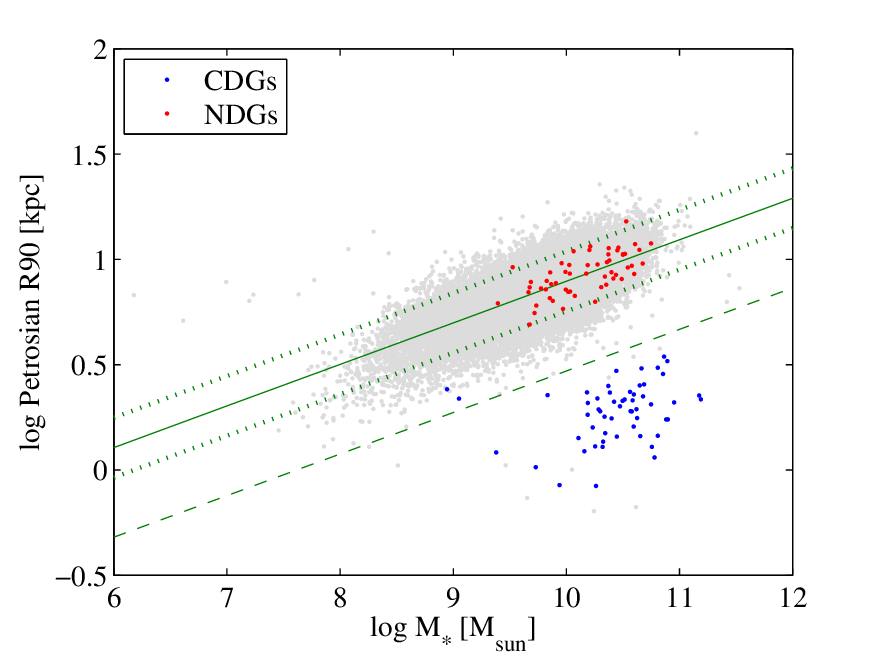}
    \caption{
    Mass--Size relation based on the \citet{2003MNRAS.341...33K} stellar masses.
    Red and blue dots denote our NDGs and CDGs, respectively.
    The solid represents the fitted size-mass relation.
    The dashed line represents $3\sigma$ below the fitted line, and the dotted lines represent one $1\sigma$ around the fitted line.}
    \label{fig9}
\end{figure}

We also investigate the possible bias on our mass estimation by comparing the results of \cite{2003MNRAS.341...33K}.
\citet{2003MNRAS.341...33K} estimated stellar masses based on the Bayesian methodology with a \citet{2001MNRAS.322..231K} initial mass function.
Figure \ref{fig9} displays the mass--size relation based on the stellar masses of \citet{2003MNRAS.341...33K}, which exhibits a feature similar to the relation shown in Figure \ref{fig1}.
Applying the same data selection used in Figure \ref{fig1} to the data in Figure \ref{fig9} would result in almost the same CDGs and NDGs, leading to the same conclusions.

\section{Summary}

We studied the spiral galaxies with compact discs and normal discs with similar mass using SDSS data.
We selected the CDGs and NDGs according to their mass-size relationship.
We determined the surface brightness profiles of the CDGs and NDGs, and then fitted to a S{\'e}rsic model both using $n=1$ and $n$ as a free parameter.
The CDGs display significantly brighter centers than the NDGs and exhibit distinct surface brightness profiles.
We determined the color gradient of the selected galaxies from their images.
The CDGs have zero color gradients whereas the NDGs have negative color gradients.
We fitted the photometry of the inner and outer parts to the \citet{2003MNRAS.344.1000B} models to obtain the stellar ages.
The CDGs have older stellar ages than the NDGs, both within and beyond their effective radii.
Although the CDGs are selected as the smallest galaxies in each stellar mass range as \citet{2021MNRAS.507..300S} selected the MCGs, the CDGs are not as small as the MCGs.
The CDGs belong to a different population.
We suggest that the CDGs are galaxies in the quenching phase after the central starburst.

\section*{Acknowledgements}

This work was supported by the Ministry of Science and Technology of Taiwan (grants MOST 111-2811-M-008-064 and 110-2112-M-008-021-MY3).

Funding for the Sloan Digital Sky Survey IV has been provided by the Alfred P. Sloan Foundation, the U.S. Department of Energy Office of Science, and the Participating Institutions. 

SDSS-IV acknowledges support and resources from the Center for High Performance Computing  at the University of Utah. The SDSS website is www.sdss.org.

SDSS-IV is managed by the Astrophysical Research Consortium for the Participating Institutions of the SDSS Collaboration including the Brazilian Participation Group, the Carnegie Institution for Science, Carnegie Mellon University, Center for Astrophysics | Harvard \& Smithsonian, the Chilean Participation Group, the French Participation Group, Instituto de Astrof\'isica de Canarias, The Johns Hopkins University, Kavli Institute for the Physics and Mathematics of the Universe (IPMU) / University of Tokyo, the Korean Participation Group, Lawrence Berkeley National Laboratory, Leibniz Institut f\"ur Astrophysik Potsdam (AIP),  Max-Planck-Institut f\"ur Astronomie (MPIA Heidelberg), Max-Planck-Institut f\"ur Astrophysik (MPA Garching), Max-Planck-Institut f\"ur Extraterrestrische Physik (MPE), National Astronomical Observatories of China, New Mexico State University, New York University, University of Notre Dame, Observat\'ario Nacional / MCTI, The Ohio State University, Pennsylvania State University, Shanghai Astronomical Observatory, United Kingdom Participation Group, Universidad Nacional Aut\'onoma de M\'exico, University of Arizona, University of Colorado Boulder, University of Oxford, University of Portsmouth, University of Utah, University of Virginia, University of Washington, University of Wisconsin, Vanderbilt University, and Yale University.

\section*{Data Availability}

The data underlying this article were accessed from the Sloan Digital Sky Survey at skyserver.sdss.org/ (dr16.Galaxy, dr16.SpecObj, dr16.galSpecExtra, dr16.galSpecIndex, dr16.zooSpec, dr16.Two- MassXSC).



\bibliographystyle{mnras}
\bibliography{reference} 

\begin{thebibliography}{}
\makeatletter
\relax
\def\mn@urlcharsother{\let\do\@makeother \do\$\do\&\do\#\do\^\do\_\do\%\do\~}
\def\mn@doi{\begingroup\mn@urlcharsother \@ifnextchar [ {\mn@doi@}
  {\mn@doi@[]}}
\def\mn@doi@[#1]#2{\def\@tempa{#1}\ifx\@tempa\@empty \href
  {http://dx.doi.org/#2} {doi:#2}\else \href {http://dx.doi.org/#2} {#1}\fi
  \endgroup}
\def\mn@eprint#1#2{\mn@eprint@#1:#2::\@nil}
\def\mn@eprint@arXiv#1{\href {http://arxiv.org/abs/#1} {{\tt arXiv:#1}}}
\def\mn@eprint@dblp#1{\href {http://dblp.uni-trier.de/rec/bibtex/#1.xml}
  {dblp:#1}}
\def\mn@eprint@#1:#2:#3:#4\@nil{\def\@tempa {#1}\def\@tempb {#2}\def\@tempc
  {#3}\ifx \@tempc \@empty \let \@tempc \@tempb \let \@tempb \@tempa \fi \ifx
  \@tempb \@empty \def\@tempb {arXiv}\fi \@ifundefined
  {mn@eprint@\@tempb}{\@tempb:\@tempc}{\expandafter \expandafter \csname
  mn@eprint@\@tempb\endcsname \expandafter{\@tempc}}}

\bibitem[\protect\citeauthoryear{{Ahumada} et~al.,}{{Ahumada}
  et~al.}{2020}]{2020ApJS..249....3A}
{Ahumada} R.,  et~al., 2020, \mn@doi [\apjs] {10.3847/1538-4365/ab929e}, \href
  {https://ui.adsabs.harvard.edu/abs/2020ApJS..249....3A} {249, 3}

\bibitem[\protect\citeauthoryear{{Blanton} et~al.,}{{Blanton}
  et~al.}{2005}]{2005AJ....129.2562B}
{Blanton} M.~R.,  et~al., 2005, \mn@doi [\aj] {10.1086/429803}, \href
  {https://ui.adsabs.harvard.edu/abs/2005AJ....129.2562B} {129, 2562}

\bibitem[\protect\citeauthoryear{{Brammer} et~al.,}{{Brammer}
  et~al.}{2012}]{2012ApJS..200...13B}
{Brammer} G.~B.,  et~al., 2012, \mn@doi [\apjs] {10.1088/0067-0049/200/2/13},
  \href {https://ui.adsabs.harvard.edu/abs/2012ApJS..200...13B} {200, 13}

\bibitem[\protect\citeauthoryear{{Brinchmann}, {Charlot}, {White}, {Tremonti},
  {Kauffmann}, {Heckman}  \& {Brinkmann}}{{Brinchmann}
  et~al.}{2004}]{2004MNRAS.351.1151B}
{Brinchmann} J.,  {Charlot} S.,  {White} S.~D.~M.,  {Tremonti} C.,  {Kauffmann}
  G.,  {Heckman} T.,   {Brinkmann} J.,  2004, \mn@doi [\mnras]
  {10.1111/j.1365-2966.2004.07881.x}, \href
  {https://ui.adsabs.harvard.edu/abs/2004MNRAS.351.1151B} {351, 1151}

\bibitem[\protect\citeauthoryear{{Bruzual} \& {Charlot}}{{Bruzual} \&
  {Charlot}}{2003}]{2003MNRAS.344.1000B}
{Bruzual} G.,  {Charlot} S.,  2003, \mn@doi [\mnras]
  {10.1046/j.1365-8711.2003.06897.x}, \href
  {https://ui.adsabs.harvard.edu/abs/2003MNRAS.344.1000B} {344, 1000}

\bibitem[\protect\citeauthoryear{{Bundy} et~al.,}{{Bundy}
  et~al.}{2015}]{2015ApJ...798....7B}
{Bundy} K.,  et~al., 2015, \mn@doi [\apj] {10.1088/0004-637X/798/1/7}, \href
  {https://ui.adsabs.harvard.edu/abs/2015ApJ...798....7B} {798, 7}

\bibitem[\protect\citeauthoryear{{Chabrier}}{{Chabrier}}{2003}]{2003PASP..115..763C}
{Chabrier} G.,  2003, \mn@doi [\pasp] {10.1086/376392}, \href
  {https://ui.adsabs.harvard.edu/abs/2003PASP..115..763C} {115, 763}

\bibitem[\protect\citeauthoryear{{Du}, {Cheng}, {Wu}, {Zhu}  \& {Wang}}{{Du}
  et~al.}{2019}]{2019MNRAS.483.1754D}
{Du} W.,  {Cheng} C.,  {Wu} H.,  {Zhu} M.,   {Wang} Y.,  2019, \mn@doi [\mnras]
  {10.1093/mnras/sty2976}, \href
  {https://ui.adsabs.harvard.edu/abs/2019MNRAS.483.1754D} {483, 1754}

\bibitem[\protect\citeauthoryear{{Fathi}}{{Fathi}}{2010}]{2010ApJ...722L.120F}
{Fathi} K.,  2010, \mn@doi [\apjl] {10.1088/2041-8205/722/1/L120}, \href
  {https://ui.adsabs.harvard.edu/abs/2010ApJ...722L.120F} {722, L120}

\bibitem[\protect\citeauthoryear{{Fathi}, {Allen}, {Boch}, {Hatziminaoglou}  \&
  {Peletier}}{{Fathi} et~al.}{2010}]{2010MNRAS.406.1595F}
{Fathi} K.,  {Allen} M.,  {Boch} T.,  {Hatziminaoglou} E.,   {Peletier} R.~F.,
  2010, \mn@doi [\mnras] {10.1111/j.1365-2966.2010.16812.x}, \href
  {https://ui.adsabs.harvard.edu/abs/2010MNRAS.406.1595F} {406, 1595}

\bibitem[\protect\citeauthoryear{{Fern{\'a}ndez Lorenzo}, {Sulentic},
  {Verdes-Montenegro}  \& {Argudo-Fern{\'a}ndez}}{{Fern{\'a}ndez Lorenzo}
  et~al.}{2013}]{2013MNRAS.434..325F}
{Fern{\'a}ndez Lorenzo} M.,  {Sulentic} J.,  {Verdes-Montenegro} L.,
  {Argudo-Fern{\'a}ndez} M.,  2013, \mn@doi [\mnras] {10.1093/mnras/stt1020},
  \href {https://ui.adsabs.harvard.edu/abs/2013MNRAS.434..325F} {434, 325}

\bibitem[\protect\citeauthoryear{{Ferr{\'e}-Mateu}, {Vazdekis}, {Trujillo},
  {S{\'a}nchez-Bl{\'a}zquez}, {Ricciardelli}  \& {de la
  Rosa}}{{Ferr{\'e}-Mateu} et~al.}{2012}]{2012MNRAS.423..632F}
{Ferr{\'e}-Mateu} A.,  {Vazdekis} A.,  {Trujillo} I.,
  {S{\'a}nchez-Bl{\'a}zquez} P.,  {Ricciardelli} E.,   {de la Rosa} I.~G.,
  2012, \mn@doi [\mnras] {10.1111/j.1365-2966.2012.20897.x}, \href
  {https://ui.adsabs.harvard.edu/abs/2012MNRAS.423..632F} {423, 632}

\bibitem[\protect\citeauthoryear{{Freeman}}{{Freeman}}{1970}]{1970ApJ...160..811F}
{Freeman} K.~C.,  1970, \mn@doi [\apj] {10.1086/150474}, \href
  {https://ui.adsabs.harvard.edu/abs/1970ApJ...160..811F} {160, 811}

\bibitem[\protect\citeauthoryear{{Galaz}, {Herrera-Camus}, {Garcia-Lambas}  \&
  {Padilla}}{{Galaz} et~al.}{2011}]{2011ApJ...728...74G}
{Galaz} G.,  {Herrera-Camus} R.,  {Garcia-Lambas} D.,   {Padilla} N.,  2011,
  \mn@doi [\apj] {10.1088/0004-637X/728/2/74}, \href
  {https://ui.adsabs.harvard.edu/abs/2011ApJ...728...74G} {728, 74}

\bibitem[\protect\citeauthoryear{{Graham} \& {Driver}}{{Graham} \&
  {Driver}}{2005}]{2005PASA...22..118G}
{Graham} A.~W.,  {Driver} S.~P.,  2005, \mn@doi [\pasa] {10.1071/AS05001},
  \href {https://ui.adsabs.harvard.edu/abs/2005PASA...22..118G} {22, 118}

\bibitem[\protect\citeauthoryear{{Haines}, {McIntosh}, {S{\'a}nchez},
  {Tremonti}  \& {Rudnick}}{{Haines} et~al.}{2015}]{2015MNRAS.451..433H}
{Haines} T.,  {McIntosh} D.~H.,  {S{\'a}nchez} S.~F.,  {Tremonti} C.,
  {Rudnick} G.,  2015, \mn@doi [\mnras] {10.1093/mnras/stv989}, \href
  {https://ui.adsabs.harvard.edu/abs/2015MNRAS.451..433H} {451, 433}

\bibitem[\protect\citeauthoryear{{Kauffmann} et~al.,}{{Kauffmann}
  et~al.}{2003}]{2003MNRAS.341...33K}
{Kauffmann} G.,  et~al., 2003, \mn@doi [\mnras]
  {10.1046/j.1365-8711.2003.06291.x}, \href
  {https://ui.adsabs.harvard.edu/abs/2003MNRAS.341...33K} {341, 33}

\bibitem[\protect\citeauthoryear{{Kroupa}}{{Kroupa}}{2001}]{2001MNRAS.322..231K}
{Kroupa} P.,  2001, \mn@doi [\mnras] {10.1046/j.1365-8711.2001.04022.x}, \href
  {https://ui.adsabs.harvard.edu/abs/2001MNRAS.322..231K} {322, 231}

\bibitem[\protect\citeauthoryear{{Lintott} et~al.,}{{Lintott}
  et~al.}{2008}]{2008MNRAS.389.1179L}
{Lintott} C.~J.,  et~al., 2008, \mn@doi [\mnras]
  {10.1111/j.1365-2966.2008.13689.x}, \href
  {https://ui.adsabs.harvard.edu/abs/2008MNRAS.389.1179L} {389, 1179}

\bibitem[\protect\citeauthoryear{{Lintott} et~al.,}{{Lintott}
  et~al.}{2011}]{2011MNRAS.410..166L}
{Lintott} C.,  et~al., 2011, \mn@doi [\mnras]
  {10.1111/j.1365-2966.2010.17432.x}, \href
  {https://ui.adsabs.harvard.edu/abs/2011MNRAS.410..166L} {410, 166}

\bibitem[\protect\citeauthoryear{{Monnier Ragaigne}, {van Driel}, {Schneider},
  {Jarrett}  \& {Balkowski}}{{Monnier Ragaigne}
  et~al.}{2003}]{2003A&A...405...99M}
{Monnier Ragaigne} D.,  {van Driel} W.,  {Schneider} S.~E.,  {Jarrett} T.~H.,
  {Balkowski} C.,  2003, \mn@doi [\aap] {10.1051/0004-6361:20030585}, \href
  {https://ui.adsabs.harvard.edu/abs/2003A&A...405...99M} {405, 99}

\bibitem[\protect\citeauthoryear{{Schnorr-M{\"u}ller}
  et~al.,}{{Schnorr-M{\"u}ller} et~al.}{2021}]{2021MNRAS.507..300S}
{Schnorr-M{\"u}ller} A.,  et~al., 2021, \mn@doi [\mnras]
  {10.1093/mnras/stab2116}, \href
  {https://ui.adsabs.harvard.edu/abs/2021MNRAS.507..300S} {507, 300}

\bibitem[\protect\citeauthoryear{{Wu}}{{Wu}}{2018}]{2018MNRAS.473.5468W}
{Wu} P.-F.,  2018, \mn@doi [\mnras] {10.1093/mnras/stx2745}, \href
  {https://ui.adsabs.harvard.edu/abs/2018MNRAS.473.5468W} {473, 5468}

\bibitem[\protect\citeauthoryear{{Wu} et~al.,}{{Wu}
  et~al.}{2018}]{2018ApJ...868...37W}
{Wu} P.-F.,  et~al., 2018, \mn@doi [\apj] {10.3847/1538-4357/aae822}, \href
  {https://ui.adsabs.harvard.edu/abs/2018ApJ...868...37W} {868, 37}

\bibitem[\protect\citeauthoryear{{van Dokkum} et~al.,}{{van Dokkum}
  et~al.}{2011}]{2011ApJ...743L..15V}
{van Dokkum} P.~G.,  et~al., 2011, \mn@doi [\apjl]
  {10.1088/2041-8205/743/1/L15}, \href
  {https://ui.adsabs.harvard.edu/abs/2011ApJ...743L..15V} {743, L15}

\bibitem[\protect\citeauthoryear{{van Dokkum} et~al.,}{{van Dokkum}
  et~al.}{2015}]{2015ApJ...813...23V}
{van Dokkum} P.~G.,  et~al., 2015, \mn@doi [\apj] {10.1088/0004-637X/813/1/23},
  \href {https://ui.adsabs.harvard.edu/abs/2015ApJ...813...23V} {813, 23}

\bibitem[\protect\citeauthoryear{{van der Wel} et~al.,}{{van der Wel}
  et~al.}{2014}]{2014ApJ...788...28V}
{van der Wel} A.,  et~al., 2014, \mn@doi [\apj] {10.1088/0004-637X/788/1/28},
  \href {https://ui.adsabs.harvard.edu/abs/2014ApJ...788...28V} {788, 28}

\makeatother
\end{thebibliography}




\appendix

\section{CDG data and measurements}

\begin{table*}
\centering
\caption{CDG data and measurements}
\label{tab:table_data}
\begin{tabular}{lccccccccc} 
\hline
IAU name	&	RA	&	DEC	&	redshift	&	$\mathrm{PetroR90}$	&	$\log M_*$	&	$\mu_0$	&	$R_e$	&	$\log \mathrm{age_*}$	&	$\log \mathrm{age_*}$	\\
& & & & & & & & $(R<R_e)$ & $(R>R_e)$ \\
&	degree	&	degree	&		&	arcsec	&	$\mathrm{M_\odot}$	&		mag/$\mathrm{arcsec^2}$				&		arcsec				&	yr	&	yr	\\
(1) & (2) & (3) & (4) & (5) & (6) & (7) & (8) & (9) & (10) \\
\hline
J084648.77+001812.6	&	131.7032	&	0.3035	&	0.029	&	1.45	&	9.15	&	$	20.20	\pm	0.15	$	&	$	8.98	\pm	0.25	$	&	9.94	&	9.51	\\
J101346.81-005451.3	&	153.4450	&	-0.9143	&	0.042	&	1.73	&	10.16	&	$	18.80	\pm	0.05	$	&	$	7.14	\pm	0.14	$	&	10.13	&	10.13	\\
J121300.61-005618.9	&	183.2526	&	-0.9386	&	0.075	&	2.31	&	10.43	&	$	19.71	\pm	0.16	$	&	$	5.11	\pm	0.09	$	&	10.13	&	10.13	\\
J130825.50+680851.3	&	197.1063	&	68.1476	&	0.058	&	1.83	&	9.97	&	$	20.29	\pm	0.20	$	&	$	8.78	\pm	0.17	$	&	9.54	&	10.13	\\
J083128.81+463728.9	&	127.8700	&	46.6247	&	0.048	&	2.02	&	10.30	&	$	19.24	\pm	0.20	$	&	$	3.93	\pm	0.05	$	&	9.88	&	10.13	\\
J101857.06+013936.0	&	154.7378	&	1.6600	&	0.046	&	1.61	&	9.79	&	$	19.46	\pm	0.17	$	&	$	6.96	\pm	0.03	$	&	10.02	&	9.70	\\
J151122.57+014258.3	&	227.8441	&	1.7162	&	0.038	&	1.81	&	9.66	&	$	19.88	\pm	0.18	$	&	$	5.51	\pm	0.06	$	&	10.13	&	10.13	\\
J090656.87+031051.4	&	136.7370	&	3.1810	&	0.027	&	1.86	&	8.68	&	$	20.43	\pm	0.14	$	&	$	10.31	\pm	0.11	$	&	8.86	&	8.86	\\
J161333.98+455938.8	&	243.3916	&	45.9941	&	0.050	&	2.37	&	9.78	&	$	19.17	\pm	0.20	$	&	$	4.08	\pm	0.09	$	&	8.86	&	10.13	\\
J174243.00+534951.8	&	265.6792	&	53.8311	&	0.051	&	2.14	&	10.05	&	$	19.52	\pm	0.23	$	&	$	4.70	\pm	0.05	$	&	9.98	&	9.98	\\
J170703.63+312525.3	&	256.7652	&	31.4237	&	0.031	&	2.07	&	9.88	&	$	19.83	\pm	0.29	$	&	$	10.89	\pm	0.34	$	&	10.13	&	10.13	\\
J234651.93+145209.3	&	356.7164	&	14.8693	&	0.058	&	1.87	&	9.21	&	$	20.07	\pm	0.09	$	&	$	5.03	\pm	0.05	$	&	9.16	&	10.07	\\
J120957.55+540108.4	&	182.4898	&	54.0190	&	0.050	&	2.35	&	9.99	&	$	19.01	\pm	0.13	$	&	$	3.70	\pm	0.03	$	&	9.54	&	9.54	\\
J104427.55+585411.3	&	161.1148	&	58.9032	&	0.031	&	1.94	&	9.21	&	$	19.66	\pm	0.24	$	&	$	5.53	\pm	0.07	$	&	9.48	&	9.89	\\
J102701.75+092247.6	&	156.7573	&	9.3799	&	0.046	&	2.13	&	10.00	&	$	19.82	\pm	0.16	$	&	$	5.49	\pm	0.24	$	&	9.41	&	9.41	\\
J080352.99+263123.3	&	120.9708	&	26.5232	&	0.046	&	1.75	&	9.45	&	$	19.32	\pm	0.05	$	&	$	4.52	\pm	0.03	$	&	9.95	&	10.13	\\
J141659.82+531427.5	&	214.2493	&	53.2410	&	0.075	&	2.44	&	10.70	&	$	19.36	\pm	0.33	$	&	$	3.51	\pm	0.06	$	&	10.13	&	10.13	\\
J093855.79+104613.3	&	144.7325	&	10.7704	&	0.064	&	1.73	&	10.19	&	$	19.96	\pm	0.13	$	&	$	5.22	\pm	0.14	$	&	10.13	&	10.13	\\
J155547.64+423626.0	&	238.9485	&	42.6072	&	0.046	&	2.81	&	10.57	&	$	19.06	\pm	0.17	$	&	$	4.88	\pm	0.06	$	&	10.07	&	10.07	\\
J152058.44+484013.0	&	230.2435	&	48.6703	&	0.070	&	2.29	&	10.39	&	$	19.71	\pm	0.26	$	&	$	3.95	\pm	0.05	$	&	9.99	&	9.99	\\
J121819.17+473504.5	&	184.5799	&	47.5846	&	0.066	&	1.83	&	10.14	&	$	19.66	\pm	0.07	$	&	$	3.44	\pm	0.10	$	&	9.97	&	10.13	\\
J154202.23+081835.8	&	235.5093	&	8.3100	&	0.041	&	2.50	&	10.33	&	$	19.71	\pm	0.25	$	&	$	5.33	\pm	0.05	$	&	10.13	&	10.13	\\
J095554.72+371340.8	&	148.9780	&	37.2280	&	0.040	&	1.81	&	9.70	&	$	20.12	\pm	0.16	$	&	$	6.01	\pm	0.08	$	&	9.98	&	9.98	\\
J131124.82+431232.6	&	197.8534	&	43.2091	&	0.030	&	2.35	&	9.79	&	$	19.04	\pm	0.06	$	&	$	6.03	\pm	0.09	$	&	9.72	&	9.72	\\
J142408.02+464758.3	&	216.0334	&	46.7995	&	0.013	&	3.90	&	7.64	&	$	21.39	\pm	0.02	$	&	$	9.68	\pm	0.64	$	&	9.54	&	9.06	\\
J131214.85+113532.7	&	198.0619	&	11.5924	&	0.032	&	2.86	&	9.58	&	$	20.15	\pm	0.13	$	&	$	7.03	\pm	0.13	$	&	10.13	&	10.13	\\
J110901.03+442442.1	&	167.2543	&	44.4117	&	0.061	&	1.90	&	9.50	&	$	20.00	\pm	0.19	$	&	$	7.92	\pm	0.57	$	&	9.26	&	9.99	\\
J140245.35+380357.6	&	210.6890	&	38.0660	&	0.064	&	1.75	&	10.34	&	$	19.86	\pm	0.24	$	&	$	9.75	\pm	0.04	$	&	10.13	&	10.13	\\
J122033.64+092755.1	&	185.1402	&	9.4653	&	0.025	&	4.77	&	9.55	&	$	19.90	\pm	0.16	$	&	$	2.92	\pm	0.05	$	&	10.13	&	10.13	\\
J113801.74+425543.8	&	174.5073	&	42.9288	&	0.061	&	1.87	&	9.47	&	$	20.11	\pm	0.13	$	&	$	5.47	\pm	0.09	$	&	9.70	&	9.34	\\
J152845.27+072134.2	&	232.1887	&	7.3595	&	0.042	&	1.38	&	9.53	&	$	19.60	\pm	0.21	$	&	$	7.93	\pm	0.07	$	&	9.86	&	10.13	\\
J164227.92+265848.4	&	250.6164	&	26.9801	&	0.068	&	1.73	&	9.35	&	$	19.90	\pm	0.19	$	&	$	5.39	\pm	0.12	$	&	8.81	&	8.81	\\
J170525.98+221617.9	&	256.3583	&	22.2716	&	0.048	&	1.83	&	10.24	&	$	19.48	\pm	0.18	$	&	$	5.04	\pm	0.06	$	&	10.13	&	10.13	\\
J160051.60+280623.6	&	240.2150	&	28.1066	&	0.035	&	2.28	&	10.11	&	$	18.98	\pm	0.17	$	&	$	5.92	\pm	0.32	$	&	9.70	&	10.13	\\
J153235.59+300558.2	&	233.1483	&	30.0995	&	0.067	&	1.65	&	9.95	&	$	19.40	\pm	0.13	$	&	$	3.68	\pm	0.09	$	&	9.34	&	9.34	\\
J151659.23+423553.7	&	229.2468	&	42.5983	&	0.040	&	1.62	&	9.37	&	$	19.40	\pm	0.15	$	&	$	5.79	\pm	0.13	$	&	9.68	&	9.72	\\
J160119.97+085009.3	&	240.3332	&	8.8359	&	0.017	&	2.47	&	9.61	&	$	19.34	\pm	0.15	$	&	$	12.29	\pm	0.13	$	&	10.13	&	10.13	\\
J161303.59+300413.7	&	243.2650	&	30.0705	&	0.054	&	1.69	&	9.65	&	$	20.33	\pm	0.15	$	&	$	6.59	\pm	0.23	$	&	9.48	&	9.48	\\
J160154.01+315331.3	&	240.4751	&	31.8920	&	0.045	&	2.37	&	10.67	&	$	19.39	\pm	0.29	$	&	$	6.70	\pm	0.18	$	&	10.13	&	10.13	\\
J161902.36+214823.3	&	244.7598	&	21.8065	&	0.038	&	1.73	&	9.39	&	$	19.05	\pm	0.09	$	&	$	4.92	\pm	0.04	$	&	9.01	&	9.23	\\
J092218.40+651907.4	&	140.5767	&	65.3187	&	0.038	&	3.97	&	10.26	&	$	20.02	\pm	0.13	$	&	$	6.20	\pm	0.16	$	&	10.13	&	10.13	\\
J102231.84+363514.1	&	155.6327	&	36.5873	&	0.026	&	2.00	&	8.52	&	$	19.65	\pm	0.05	$	&	$	4.16	\pm	0.04	$	&	8.86	&	9.16	\\
J093130.72+262801.6	&	142.8780	&	26.4671	&	0.065	&	2.00	&	9.94	&	$	19.55	\pm	0.11	$	&	$	5.14	\pm	0.13	$	&	9.36	&	10.13	\\
J104109.43+344301.0	&	160.2893	&	34.7170	&	0.059	&	2.52	&	9.99	&	$	18.88	\pm	0.13	$	&	$	5.11	\pm	0.03	$	&	8.71	&	8.71	\\
J143905.91+230258.6	&	219.7747	&	23.0496	&	0.066	&	1.85	&	9.86	&	$	19.73	\pm	0.35	$	&	$	4.42	\pm	0.07	$	&	9.01	&	10.13	\\
J152132.68+182643.5	&	230.3862	&	18.4454	&	0.057	&	1.75	&	9.06	&	$	19.59	\pm	0.05	$	&	$	4.01	\pm	0.08	$	&	10.13	&	9.90	\\
J151415.45+203322.0	&	228.5644	&	20.5561	&	0.039	&	1.61	&	9.74	&	$	19.16	\pm	0.04	$	&	$	5.08	\pm	0.06	$	&	9.81	&	9.94	\\
J135120.57+244221.1	&	207.8357	&	24.7059	&	0.065	&	1.80	&	10.50	&	$	19.43	\pm	0.17	$	&	$	8.32	\pm	0.45	$	&	9.70	&	9.70	\\
J135843.25+244106.6	&	209.6802	&	24.6852	&	0.075	&	1.79	&	10.43	&	$	19.64	\pm	0.08	$	&	$	4.51	\pm	0.12	$	&	10.06	&	10.06	\\
J132350.40+253342.8	&	200.9600	&	25.5619	&	0.064	&	1.44	&	9.36	&	$	19.47	\pm	0.20	$	&	$	4.71	\pm	0.07	$	&	8.81	&	9.60	\\
J162156.68+144817.2	&	245.4862	&	14.8048	&	0.029	&	1.76	&	9.06	&	$	19.62	\pm	0.07	$	&	$	5.58	\pm	0.08	$	&	10.10	&	10.10	\\
J114410.44+294527.3	&	176.0435	&	29.7576	&	0.046	&	1.94	&	10.07	&	$	19.96	\pm	0.29	$	&	$	8.33	\pm	0.28	$	&	10.01	&	10.01	\\
J111046.45+284133.8	&	167.6936	&	28.6927	&	0.033	&	2.30	&	9.99	&	$	18.74	\pm	0.18	$	&	$	3.85	\pm	0.04	$	&	9.92	&	10.13	\\
J103702.74+202553.1	&	159.2614	&	20.4314	&	0.043	&	2.04	&	10.09	&	$	19.28	\pm	0.25	$	&	$	6.01	\pm	0.11	$	&	10.02	&	10.02	\\
J124726.14+294715.8	&	191.8590	&	29.7877	&	0.022	&	1.47	&	8.77	&	$	20.36	\pm	0.18	$	&	$	16.32	\pm	0.12	$	&	10.04	&	10.04	\\
J105952.49+251633.4	&	164.9687	&	25.2760	&	0.021	&	1.74	&	8.60	&	$	19.51	\pm	0.10	$	&	$	6.19	\pm	0.04	$	&	9.40	&	10.13	\\
J114642.56+235748.1	&	176.6773	&	23.9634	&	0.021	&	1.48	&	8.48	&	$	20.23	\pm	0.05	$	&	$	17.78	\pm	0.26	$	&	9.11	&	10.08	\\
J103138.87+192705.2	&	157.9120	&	19.4515	&	0.046	&	2.38	&	9.64	&	$	19.45	\pm	0.10	$	&	$	6.11	\pm	0.03	$	&	9.32	&	9.32	\\
\end{tabular}
\end{table*}

\begin{table*}
\centering
\contcaption{CDG data and measurements}
\begin{tabular}{lccccccccc} 
\hline
IAU name	&	RA	&	DEC	&	redshift	&	$\mathrm{PetroR90}$	&	$\log M_*$	&	$\mu_0$	&	$R_e$	&	$\log \mathrm{age_*}$	&	$\log \mathrm{age_*}$	\\
& & & & & & & & $(R<R_e)$ & $(R>R_e)$ \\
&	degree	&	degree	&		&	arcsec	&	$\mathrm{M_\odot}$	&		mag/$\mathrm{arcsec^2}$				&		arcsec				&	yr	&	yr	\\
(1) & (2) & (3) & (4) & (5) & (6) & (7) & (8) & (9) & (10) \\
\hline
J102247.38+194730.1	&	155.6974	&	19.7917	&	0.039	&	2.50	&	9.84	&	$	19.41	\pm	0.24	$	&	$	5.44	\pm	0.07	$	&	9.63	&	10.13	\\
J101757.83+142310.4	&	154.4910	&	14.3862	&	0.047	&	2.06	&	9.95	&	$	20.10	\pm	0.16	$	&	$	5.18	\pm	0.18	$	&	9.99	&	9.99	\\
J083623.97+101250.0	&	129.0999	&	10.2139	&	0.031	&	3.57	&	9.88	&	$	20.00	\pm	0.06	$	&	$	6.37	\pm	0.12	$	&	10.13	&	10.13	\\
J114136.83-022720.3	&	175.4035	&	-2.4557	&	0.066	&	2.40	&	10.49	&	$	19.67	\pm	0.16	$	&	$	4.57	\pm	0.13	$	&	10.06	&	10.06	\\
\hline
\multicolumn{10}{l}{ Notes. (1) Name of the galaxy. (2) RA and (3) DEC in degree. (4) Spectrum redshift. (5) Petrosian $90\%$ radius of $r^{\prime}-$band in arcsec. }\\
\multicolumn{10}{l}{ (6) Logarithmic stellar mass in solar mass. (7) Central surface brightness of $r^{\prime}-$band. (8) Effective Radius of $r^{\prime}-$band in arcsec.}\\
\multicolumn{10}{l}{ (9) Logarithmic stellar age inside $R_e$. (10) Logarithmic stellar age outside $R_e$.}
\end{tabular}
\end{table*}


\bsp	
\label{lastpage}
\end{document}